\documentclass[12pt,a4paper]{article}

\usepackage{amssymb}
\usepackage{amsmath}
\usepackage{amsfonts}

\usepackage{a4wide}
\usepackage{graphicx}
\usepackage{xypic}

\newtheorem{prop}{Proposition}

\begin{document}

\title{Quadratic algebras for three dimensional non degenerate superintegrable systems with quadratic integrals of motion\footnote{Detailed version of the talk given at the XXVII Colloquium on Group Theoretical Methods in Physics, Yerevan, Armenia, Aug. 2008}}

\author{ Y. Tanoudis\thanks{tanoudis@math.auth.gr} and C. Daskaloyannis\thanks{daskalo@math.auth.gr} \\
Mathematics Department\\
Aristotle University of Thessaloniki\\
54124  Thessaloniki- Greece}
\date{Jan 2009}

\maketitle

\begin{abstract}
 The three dimensional superintegrable systems with quadratic integrals of motion have five functionally independent integrals, one among them is the Hamiltonian. Kalnins, Kress and Miller have proved that in the case of non degenerate potentials there is a sixth quadratic integral, which is linearly independent of the other integrals.  The existence of this sixth integral imply that the  integrals of motion form  a ternary {parafermionic-like} quadratic Poisson algebra with five generators. We show that in all the non degenerate cases (with one exception) there are at least two subalgebras of three integrals having a Poisson quadratic algebra structure, which is similar to the two dimensional case.
\end{abstract}

\section{Introduction}\label{int1}

In classical mechanics, a superintegrable or completely integrable
is a Hamiltonian system with a maximum number of integrals. Two well
known examples are the harmonic oscillator and the Coulomb
potential. In the $N$-dimensi\-onal space the superintegrable system
has $2 N-1$ integrals, one among them is the Hamiltonian.

  A compilation of the known three dimensional superintegrable potentials with quadratic integrals and their integrals with quadratic integrals of motion can be found  in \cite{Evans90}.  Kalnins, Kress and Miller\cite{KalKrMi07,KalKrMi07JPA} studying the Darboux equations in three dimensions have classified the complex systems. In fact Kalnins, Kress and Miller studied non degenerate potentials (depending on 4 parameters)\cite{KalKrMi07} and the degenerate potentials (depending on 3 parameters)\cite{KalKrMi07JPA}. The real systems, i.e the systems possessing a real potential can be found in the Evans seminal paper \cite{Evans90}.  One of the  results of the paper \cite{KalKrMi07} is  the so called "5 to 6" Theorem, which states that any three dimensional non degenarate superintegrable system with quadratic integrals of motion has always a sixth integral $F$ that is linearly independent but not functionally dependent regarding the set of five integrals $A_1,A_2,B_1,B_2,H $. This statement leads to the result that any three dimensional superintegrable non degenarate system form a parafermionic-like Poisson algebra of special character, whose the definition is given in Section \ref{para}.
In Section \ref{threedim} we show that the special character means that the structure of this algebra is such that contains one at least subalgebra.
%The number of the containing subalgebras depending from the extra symmetries who characterized %the potential.
 %According to the last statement we have so far, a general class that all three dimensional %superintegrable systems obey and containing two subalgebras.
 %The extra symmetries lead us to two more classes Class I, II that contains three and one extra %subalgebras respectively.

%The non degenerate potentials have also another structure. The $C_1=\{ A_1, B_1\}_P$, $C_2=\{ %A_2, B_2\}_P$, $D=\{B_1,B_2 \}_P$ form an algebra
%that expand in terms of $C_1,C_2,D$ and coefficients any linear combination of integrals $ %A_1,A_2,B_1,B_2, H,F$

The degenerate superintegrable systems lead to non linear deviations of the quadratic algebras. Two between them are special cases of non degenerate potentials, which have a sixth integral of motion but their algebras are not quadratic ones. The algebra of the degenerate systems is under current investigation.

\section{Parafermionic-like Poisson algebras}\label{para}

The Universal Enveloping Algebra  $U(\mathfrak{g})$ of the Lie algebra  $\mathfrak{g}$ with generators $x_1,x_2,\ldots,x_n$
satisfy the relations
\[
[x_i,x_j] =x_i x_j - x_j x_i= \sum\limits_{m}c_{ij}^m x_m
\]
The generators satisfy the obvious ternary (trilinear) relations
\begin{equation}\label{eq:DefLieTriple}
T\left(x_i,x_j,x_k\right) \mathop{\equiv}\limits^{\rm def } [x_i,[x_j,x_k]]= \sum\limits_{n} d_{i;jk}^n x_n, \quad
\mbox{ where } \; d_{i;jk}^n = \sum\limits_{m} c_{i m}^n c_{jk}^m
\end{equation}
Generally a ternary algebra is an associative algebra $\mathcal{A}$ satisfying whose the generators satisfy relations like the following one
\[
T\left(x_i,x_j,x_k\right) = \sum\limits_{n} d_{i;jk}^n x_n
\]
where $T\, :\, \mathcal{A}\otimes\mathcal{A}\otimes\mathcal{A} \longrightarrow \mathcal{A}$ is a trilinear map. If this trilinear map is defined as in eq. (\ref{eq:DefLieTriple}) the corresponding algebra is an example of the triple Lie algebras, which were introduced by Jacobson \cite{Jac51} in 1951. At the same time Green \cite{Green53} was introduced the
parafermionic algebra as an associative algebra, whose operators $f_i^\dagger, f_i  $  satisfy the ternary relations:
\[
\begin{array}{c}
\left[ \, f_k,  \,  \left[ f_\ell^\dagger  ,\, f_m \right] \right] =  2 \delta_{k \ell} f_m
\\
\left[ \,  f_k,  \,  \left[ f_\ell^\dagger  ,\, f_m^\dagger   \right] \right] =
 2 \delta_{k \ell} f_m^\dagger - 2 \delta_{k m} f_\ell ^\dagger
\\
\left[ \,  f_k,  \,  \left[ f_\ell  ,\, f_m \right] \right] = 0
\end{array}
\]

We call  parafermionic Poisson algebra the  Poisson algebra satisfying
the ternary relations:
\[
\left\{ x_i, \left\{ x_j, x_k \right\}_P \right\}_P
=\sum\limits_{m} c_{i;jk}^m x_m
\]
which is the classical Poisson analogue of the Lie triple algebra (\ref{eq:DefLieTriple}).

The quadratic parafermionic Poisson algebra is a Poisson algebra satisfying
the relations:
\[
\left\{ x_i, \left\{ x_j, x_k\right\}_P \right\}_P =
\sum\limits_{m,n}  d_{i;jk}^{mn} x_m x_n
 +\sum\limits_{m} c_{i;jk}^m x_m
\]

A classical  superintegrable system with quadratic integrals of motion on a two dimensional manifold possesses has two functionally independent integrals of motion $A$ and $B$, which are in involution with the Hamiltonian $H$ of the system:
\[
\left\{H,\, A\right\}_P=0, \quad \left\{H,\, B\right\}_P=0
\]
the Poisson bracket $  \left\{A,\, B\right\}_P$ is different to zero and it is generally an integral of motion cubic in momenta, therefore it could not be in general  a linear combination of the integrals $H,\, A,\, B$.
Generally if we the Poisson brackets of the integrals of motion
 $\left\{ A,\left\{A,\,B\right\}_P\right\}_P$, $\left\{\left\{A,\,B\right\}_P ,B\right\}_P$ are not linear functions of the intergals of motion, therefore they don't close in a Lie Poisson algebra with three generators. If we consider all the nested Poisson brackets of the integrals of motion, generally they don't close in an Poisson Lie algebra structure.

All the known two dimensional superintegrable systems with quadratic integrals of motion the have a common structure \cite{Das00,Das01,DasYps06,DasTan07}:
\begin{equation}
\label{eq:twodim}
\begin{array}{c}
\left\{H,\, A\right\}_P=0, \quad \left\{H,\, B\right\}_P=0,\quad
\left\{A,\, B\right\}_P\ne 0
\left\{ A, B\right\}_{P}^2= 2 F(A,H,B)\\
\left\{ A , \left\{A,B\right\}_P\, \right\}_P = \dfrac{\partial F}{ \partial B} \;  , \;
\left\{ B ,  \left\{A,B\right\}_P\, \right\}_P =-\dfrac{\partial F}{ \partial A}
\end{array}
\end{equation}
where $F=F(A,B,H)$ is a cubic function of the integrals of motion
\begin{equation}\label{eq:structure_two_dim}
\begin{array}{rl}
F(A,B,H)=&\alpha A^3 + \beta B^3 +  \gamma A^2 B + \delta A B^2+
 \left(\epsilon_0 + \epsilon_1 H \right)  A^2  +
 \left(\zeta_0 + \zeta_1 H\right)  B^2  +\\
+&
\left(\eta_0+ \eta_1 H \right) A B  +
 \left(\theta_0 + \theta_1 H + \theta_2 H^2\right)  A+\\
+&\left( \kappa_0 + \kappa_1 H + \kappa_2 H^2\right) B + \left(\lambda_0+ \lambda_1 H + \lambda_2 H^2 + \lambda_3 H^3\right)
\end{array}
\end{equation}
where the greek letters are constants.

%\begin{prop}
Therefore any two dimensional superintegrable system correspond to some
parafermionic-like quadratic Poisson Algebra with
two generators. In fact the two dimensional systems can be classified in six classes by classifying the corresponding parafermionic-like quadratic Poisson algebras \cite{DasYps06}. By classifying the correponding associative algebras all the two dimensional superintegrable systems with quadratic integrals of motion are classified too \cite{DasTan07}.
%\end{prop}

\section{The structure of the Poisson algebra of three dimensional potentials}\label{threedim}

The known three-dimensional superintegrable systems defined on a flat space are described by the Hamiltonian
\begin{equation}
H=\frac{1}{2} (p_{x}^2+p_{y}^2+p_{z}^2)+V(x,y,z)
\end{equation}
were initially studied by  Evans  \cite{Evans90}. Kalnins, Kress, and Miller in \cite{KalKrMi07} have classified all the three dimensional superintegrable systems with quadratic integrals of motion. These potentials are distinguished to the non degenerate ones, which are potentials which depend on four parameters, i.e. they are linear combination of four potedntials. The degenerate potential depend on less than four parameters.
One of the general results in \cite{KalKrMi07} is the so called "5 to 6" theorem:

\indent\textbf{5$\to$6 Theorem}:
\textit{
Let $V$ be a nondegenerate potential (depending on 4 parameters) corresponding to
  a conformally flat space in 3 dimensions
  \[ ds^2= g(x,y,z) (dx^2+dy^2 +dz^2)\]
   that is superintegrable and there are 5 functionally
independent constants of the motion ${\mathcal L}=\{{ S}_\ell:\ell=1,\cdots 5\}$    There is always a $6$th quadratic integral ${S}_6$ that is
functionally dependent on $\mathcal L$, but linearly independent
}

By srudying all the known non degenerate potentials, we can prove the following theorem:

\begin{prop}
%\textbf{Proposition}
%\textit
{In the case of the non degenerate with quadratic integrals of motion, on a conformally flat
manifold, the integrals of motion satisfy a  parafermionic-like quadratic Poisson Algebra with 5  generators which described from the following:
\begin{equation}\label{eq:PoissonParaFermionicAlgebra}
\left\{ S_i,\, \left\{ S_j, \, S_k \right\}_{P} \right\}_{P}=
\sum\limits_{m n} d^{m,n}_{i;jk} S_m S_n + \sum\limits_{m} c^{m}_{i;jk} S_m
\end{equation}
}
\end{prop}

In all the three dimensional superintegrable systems with quadratic integrals with 4 or 3 parameters, the integrals of motion satisfy a "special" form of the Poisson Parafermionic-like algebra (\ref{eq:PoissonParaFermionicAlgebra}).
\begin{prop}
In all the known cases (with only one exception) the non degenerate systems we can choose beyond the Hamiltonian $H$ four functionally independent integrals of motion $A_1,\, B_1, \, A_2, \, B_2$, and one additional quadratic integral of motion $F$, such that all the integrals of motion
are linearly independent. These integrals satisfy a Poisson parafermionic-like algebra (\ref{eq:PoissonParaFermionicAlgebra}).  The "special" form of the algebra defined by the integrals  $A_1,\, B_1, \, A_2, \, B_2$ is characterized by two cubic functions
\[
F_1=F_1\left(A_1,A_2,B_1,H\right) \quad
F_2=F_2\left(A_1,A_2,B_2,H\right)
\]
and satisfy the relations:
\begin{equation}\label{eq:SpecialStructure}
\begin{array}{c}
\left\{A_1,A_2\right\}_P=\left\{A_1,B_2\right\}_P= \left\{A_2,B_1\right\}_P=0,
\\
\begin{array}{l}
\left\{A_1,B_1\right\}_P^2= 2 F_1(A_1,A_2,H,B_1)=\mbox{cubic function}
\\
\left\{A_2,B_2\right\}_P^2= 2 F_2(A_1,A_2,H,B_2)=\mbox{cubic function}
\end{array}
\\
\left\{ A_i, \left\{A_i, B_i\right\}_P \right\}_P= \dfrac{\partial F_i}{\partial B_i} \; , \; \left\{ B_i, \left\{A_i, B_i\right\}_P \right\}_P=- \dfrac{\partial F_i}{\partial A_i}
\\
\left\{ \left\{A_1, B_1\right\}_P, B_2\right\}_P= \left\{ A_1,\left\{B_1, B_2\right\}_P\right\}_P, \quad \left\{\left\{A_2, B_2\right\}_P, B_1\right\}_P=- \left\{ A_2,\left\{B_1, B_2\right\}_P\right\}_P
\end{array}
\end{equation}
\end{prop}

If we put
\[
C_1=\left\{ A_1,\, B_1\right\}_P,\quad C_2=\left\{ A_2,\, B_2\right\}_P,
\quad D=\left\{ B_1,\, B_2\right\}_P,
\]
the relations (\ref{eq:SpecialStructure}) imply the following ones:
\begin{equation}\label{eq:SpecialImplications}
\begin{array}{c}
\left\{C_1, B_2 \right\}_P C_1 - \dfrac{\partial F_1}{\partial A_2} C_2 - \dfrac{\partial F_1}{\partial B_1} D=\left\{C_2, B_1 \right\}_P C_2 - \dfrac{\partial F_2}{\partial A_1} C_1 + \dfrac{\partial F_2}{\partial B_2} D=0
\\
\left\{ C_1,C_2\right\}_P=
%\begin{split}
\dfrac{
\left|
\begin{array}{cc}
\left\{ A_1,D\right\}_P & - \dfrac{\partial F_1}{\partial A_2}\\
\dfrac{\partial F_2 }{\partial A_1}  & \left\{ A_2,D\right\}_P
\end{array}
 \right|
}{D} =
\dfrac{
\left|
\begin{array}{cc}
- \dfrac{\partial F_1}{\partial B_1}& \left\{ A_1,D\right\}_P \\
- \dfrac{\partial F_2}{\partial B_2}&\dfrac{\partial F_2 }{A_1}
\end{array}
 \right|
}{C_2}
=
\dfrac{
\left|
\begin{array}{cc}
- \dfrac{\partial F_1}{\partial A_2}&- \dfrac{\partial F_1}{\partial B_1}\\
\left\{ A_2,D\right\}_P&- \dfrac{\partial F_2}{\partial B_2}
\end{array}
 \right|
}{C_1}
%\end{split}
\end{array}
\end{equation}
and
\[
\{C_1,C_2\}_P=\frac{\displaystyle \frac{\partial F_1}{\partial B_1} \frac{\partial F_2}{\partial A_1} C_1+\frac{\partial F_1}{\partial A_2}\frac{\partial F_2}{\partial B_2} C_2+\frac{\partial F_1}{\partial B_1}\frac{\partial F_2}{\partial B_2} D}{C_1 C_2}
\]

Schematically the structure of the above algebra is described by the following "$\Pi$" shape
\begin{equation}\label{eq:Pi_diagram}
  \xymatrix{
  A_1 \ar@{--}[d] \ar@{--}[r]
                & A_2 \ar@{--}[d]  \\
  B_2 \ar@{}[ur]
                & B_1 \ar@{}[ul]   }
\end{equation}
where with dashed line represented the vanishing of Poisson bracket whereas the other brackets between the integrals are non vanishing Poisson brackets.

It is  important to notice that the integrals $A_1,B_1$ satisfy a parafermionic-like quadratic Poisson algebra similar to the algebra as in two dimensional case (\ref{eq:twodim}). The corresponding structure function to the two dimensional one  (\ref{eq:structure_two_dim}) can be written as:
\begin{equation}\label{eq:structure_three_dim}
\begin{array}{rl}
F_1(A_1,B_1,H,A_2)=&\alpha_1 A_1^3 + \beta_1 B_1^3 +  \gamma_1 A_1^2 B_1 + \delta A_1 B_1^2+
 \left(\epsilon_{01} + \epsilon_{11} H  + \epsilon_{21}A_2\right)  A_1^2  +\\
 +&\left(\zeta_{01} + \zeta_{11} H + \zeta_{21} A_2\right)  B_1^2  +
\left(\eta_{01}+ \eta_{11} H+ \eta_{21} A_2 \right) A_1 B_1  +\\
 +&\left(\theta_{01} + \theta_{11} H + \theta_{21} H^2  + \theta_{31} A_2 + \theta_{41} A_2^2 + \theta_{51} A_2 H \right)  A_1+\\
+&\left( \kappa_{01} + \kappa_{11} H + \kappa_{21} H^2+ \kappa_{31} A_2 + \kappa_{41} A_2^2 + \kappa_{51} A_2 H\right) B_1 + \\
+&\lambda_{01}+ \lambda_{11} H + \lambda_{21} H^2 + \lambda_{31} H^3+
\lambda_{41} A_2 + \lambda_{51} A_2^2 + \lambda_{61} A_2^3+ \\
+& \lambda_{71} A_2 H + \lambda_{81} A_2^2 H +  \lambda_{91} A_2 H^2
\end{array}
\end{equation}

The pair $A_2,B_2$ forms also a parafermionic-like algebra with the corresponding structure function $F_2(A_2,B_2,H,A_1)$, which has a similar form as in (\ref{eq:structure_three_dim}).

\section{Non degenerate Potentials}
In this section we give explicitly the form of the quadratic algebras for the non degenerate systems given by Kalnins, Kress and Miller (KKM)\cite{KalKrMi07}  and Evans (Ev) \cite{Evans90}. The full algebra is given after some definitions.

\begin{itemize}
\item
\underline{ KKM Potential $V_I$}
This potential which is also referred as harmonic oscillator potential  get six, known,  linearly independent integrals $H,A_{1},A_{2},B_{1},B_{2}, F$ and studied in \cite{KalKrMi07}, \cite{Evans90}.
\[
H=p_{x}^2+p_{y}^2+p_{z}^2+ \delta (x^2+y^2+ z^2)  +\frac{\alpha_{1}}{x^2}+\frac{\alpha_{2}}{y^2}+\frac{\alpha_{3}}{z^2}
\]
\[
A_{1}=p_{x}^2+ \delta x^2+ \frac{a_{1}}{x^2}
,\quad
A_{2}=p_{z}^2+\delta z^2+\frac{\alpha_{3}}{z^2}
\]
\[
B_{1}=J_{z}^2+k x^2+\frac{\alpha_{2} x^2}{y^2}+\frac{\alpha_{1} y^2}{x^2}
,\quad
B_{2}=J_{x}^2+\frac{\alpha_{3} y^2}{z^2}+\frac{\alpha_{2} z^2}{y^2}
\]
and
\[
F=J_{x}^2+J_{y}^2+J_{z}^2+\frac{\alpha_{2}(x^2+z^2)}{y^2}+\frac{\alpha_{3} x^2 (x^2+y^2)+\alpha_{1} z^2 (y^2+z^2)}{x^2 z^2}
\]
where,
\[
J_x=y p_z-z p_y \quad J_y=z p_x-x p_z \quad J_z=x p_y-y p_x \quad J^2=J_x^2+J_y^2+J_z^2
\]
The above integrals satisfy the following relations:
\[
\{ A_{1},A_{2}\}=\{A_{1},B_{2} \}=\{A_{2},B_{1}\}=\{ B_{1},F \}=\{ B_{2},F \}=0
\]
This relation corresponds to a complicated diagram, having five "$\Pi$" structures as in (\ref{eq:Pi_diagram})
\begin{equation}\label{eq:Pi_HarmOsc}
 \xymatrix{
  A_1 \ar@{--}[d] \ar@{--}[r]
                & A_2 \ar@{--}[d]  \\
  B_2 \ar@{--}[dr]
                & B_1 \ar@{--}[d] \\
                & F \ar@{}[u]  }
\end{equation}
The structure functions are given by the relations:
\[
\{A_{1},B_{1}\}=C_{1},\{ A_{2},B_{2}\}=C_{2},\{B_{1},B_{2}\}=D,
\{F,A_{1} \}=L,\{ F,A_{2} \}=M,\{ F, B_{2}\}=N
\]
\[
C_{1}^2=2 F_{1},\quad C_{2}^2=2 F_{2},\quad L^2=2 F_3,\quad M^2=2 F_4, \quad N^2=2 F_5
\]
\[
\begin{array}{l}
F_{1}=-8 (A_{1}^2 (\alpha_{2}+B_{1})+A_{1} B_{1} (A_{2}-H)+\delta B_{1}^2+\alpha_{1}( (A_{1}+A_{2}-H)^2-4 \alpha_{2}\delta ))
\\
F_{2}=-8 ( (A_{1}+A_{2}-H) (\alpha_{3} A_{1}+A_{2}(\alpha_{3}+B_{2})-\alpha_{3} H)+\delta B_{2}^2+\alpha_{2} (A_{2}^2-4 \alpha_{3} \delta) )
\\
F_3=-8 \Big((\alpha_{2}+\alpha_{3}+F) A_{1}^2+(B_{2}-F) H
   A_{1}+(B_{2}-F)^2 \delta +\\
   +\alpha_{1} \left((A_{1}-H)^2-4
   (\alpha_{2}+\alpha_{3}+B_{2}) \delta \right)\Big)
\\
F_4=-8 (\alpha_{1}+\alpha_{2}+\alpha_{3}+F) A_{2}^2+8 (2 \alpha_{3}-B_{1}+F) H
   A_{2}\\
   -8 \alpha_{3} H^2+8 \left(-(B_{1}-F)^2+4 \alpha_{1} \alpha_{3}+4
   \alpha_{2} \alpha_{3}+4 \alpha_{3} B_{1}\right) \delta
\\
F_5=8 \left(-\alpha_{2} (B_{1}+B_{2}-F)^2+\alpha_{1} \left(4 \alpha_{2}
   \alpha_{3}-B_{2}^2\right)-B_{1} (\alpha_{3} B_{1}+B_{2}
   (B_{1}+B_{2}-F))\right)
\end{array}
\]
The full algebra is given by the following relations:
\[ \{ A_{1}, \{A_1,B_1\} \}=\frac{\partial F_{1}}{\partial B_{1}}=-8 A_{1} (A_1+ A_{2} - H )-16 \delta B_{1} \]
\[ \{ A_{2},\{A_2,B_2\} \}=\frac{\partial F_{2}}{\partial B_{2}}=-8 A_{2} (A_2 + A_{1} - H )-16 \delta B_{2}\]
\[ \{\{A_{1},F\},A_{1}\}=-\frac{\partial F_3}{\partial F}=8 A_{1} (A_1-  H) -16 \delta (B_{2}-F) \]
\[ \{\{A_{2},F\},A_{2}\}=-\frac{\partial F_4}{\partial F}=8 A_{2} (A_2 -  H)-16 \delta (B_{1}-F) \]
\[ \{\{A_{1},F\},A_{2} \}=\{\{A_{2},F\},A_{1} \}=8 A_{1} A_{2}+16 \delta (B_{1}+B_{2}- F )\]
\[ \{ \{A_1,B_1\},F \}=\{\{A_{1},F\},B_{2}\}=\{\{A_{2},F\},B_{1}\}=\{\{B_1,B_2\},F\}=0 \]
\[ \{ \{A_2,B_2\},B_{1}\}=\{ \{B_1,B_2\},A_{2}\}=8 A_{1} (B_{1}-F)+ 8 (B_{1}+B_{2}-F) (A_{2}-H) \]
\[ \{ A_{1},\{B_1,B_2\} \}=\{ \{A_1,B_1\},B_{2}\}=8 A_2 (B_2-F)+8 (B_{1}+B_{2}-F) (A_{1}-H)  \]
\[ \{ B_{1}, \{B_1,B_2\}\}=\frac{\partial F_5}{\partial B_2}= -8 (B_1+2 B_2-F+2 \alpha_2) B_1-16 (\alpha_1+\alpha_2) B_2+16 \alpha_2 F\]
\[ \{ \{B_1,B_2\}, B_{2}\}=\frac{\partial F_5}{\partial B_1}= -8 (B_2+2 B_1-F+2 \alpha_2) B_2-16 (\alpha_2+\alpha_3) B_1+16 \alpha_2 F\]
\[ \{\{A_{1},F\},F\}=\frac{\partial F_3}{\partial A_1}=-16 \alpha_1 (A_1-H)-16 (\alpha_2+\alpha_3) A_1-8 (2 A_1-H) F-8 B_2 H\]
\[ \{\{A_{2},F\},F\}=\frac{\partial F_4}{\partial A_2}=-16 \alpha_3 (A_2-H)-16 (\alpha_1+\alpha_2) A_2-8 (2 A_2-H) F-8 B_1 H \]
\[ \{ \{A_1,B_1\},B_{1} \}= \frac{\partial F_{1}}{\partial A_{1}}=-16 \alpha_1 (A_1+A_2-H)-16 \alpha_2 A_1-8 (2 A_1-A_2) B_1+8 B_1 H \]
\[ \{ \{A_1,B_2\},B_{2}\}=\frac{\partial F_{2}}{\partial A_{2}}=-16 \alpha_3 (A_1+A_2-H)-16 \alpha_2 A_2-8 (2 A_2-A_1) B_2+8 B_2 H \]
\[
\begin{array}{cl}
\{\{A_1,B_1\},F\} =\{\{A_{1},F\},B_{1}\}= & -16 (\alpha_{1}+\alpha_{2}) A_{1} -16 \alpha_{1} A_{2}-8 A_{1} B_{1}+ 8 A_{2} B_{2}\\ &-8 F(A_{1}+A_{2}-H)+(16 \alpha_{1}-8 B_{2}) H
\end{array}      \]
\[
\begin{array}{cl}
\{\{A_2,B_2\},F\}=\{\{A_{2},F\},B_{2}\}= & -16 A_{2}(\alpha_{2}+\alpha_{3})-16 \alpha_{3}A_{1} -8 A_2 B_2 + 8 A_1 B_1\\ & -8 F (A_1+A_2-H)+(16 \alpha_3-8 B_1) H
\end{array}
\]
The second algebra which expand in terms of $C_{1},C_{2},D$  with coefficients any linear combination of integrals $A_{1},A_{2},B_{2},B_{2},F,H$ is:
\[
\{ C_{1},C_{2}\}=-8 A_{2} C_{1}+8 A_{1} C_{2}+16 \delta D, \{C_{1},D \}=8 (F- B_{1}- B_{2} ) C_{1}-8 (2 \alpha_{1} +B_{1}) C_{2}-8 A_{1} D
\]
\[
\{ C_{2},D \}= 8 (B_{2}+2 \alpha_{3} ) C_{1}+ 8 (B_{1}+B_{2}-F) C_{2}-8 A_{2} D
\]
The relation between  $B_{1},B_{2},C_{1}, C_{2},M,L, F$ is:
\[
C_{1}+C_{2}+M+L=0
\]
\item
\underline{ KKM Potential $V_{II}$}
This potential get six, known,  linearly independent integrals $H,A_{1},A_{2},B_{1},B_{2}, F$.  \[
H=p_{x}^2+p_{y}^2+p_{z}^2+ \alpha (x^2+y^2+ z^2)+\frac{\beta (x-i y)}{(x+i y)^3}+\frac{\gamma}{(x+i y)^2}+\frac{\delta}{z^2}
\]

\[
A_{1}=p_{z}^2+\alpha z^2+\frac{\delta}{z^2}
, \quad
A_{2}=J_{z}^2+\frac{2 i x y (\gamma -2 \beta) }{(x+i y)^2}+\frac{2 \gamma x^2}{(x+i y)^2}
\]
\[
\begin{array}{l}
B_{1}=J^2 +\\
 +\dfrac{z^2 (\beta ( 4 x y (y-i x)+z^2 (x-i y))+\gamma (x+i y)(2 x(x+i y)+z^2 ))+\delta (x- i y) (x+i y)^4}{z^2 (x+i y)^3}
\end{array}
\]
\[
B_{2}=(p_{x}+i p_{y})^2+\alpha (x+i y)^2-\frac{\beta}{(x+i y)^2}
,\quad
F=(J_{y}-i J_{x})^2+\frac{\delta (x+i y)^4-\beta z^4}{z^2 (x+i y)^2}
\]
The integrals of motion satisfy the relations
\[
\{ A_{1},A_{2}\}=\{A_{1},B_{2} \}=\{A_{2},B_{1}\}=\{ B_{1},F \}=\{ B_{2},F \}=0
\]
The corresponding "$\Pi$" shape diagram is the same as in (\ref{eq:Pi_HarmOsc}). We can define
\[
\{A_{1},B_{1}\}=C_{1},\{ A_{2},B_{2}\}=C_{2},\{B_{1},B_{2}\}=D,
 \{F,A_{1} \}=L,\{ F,A_{2} \}=M,\{ F, B_{2}\}=N
\]
\[
C_{1}^2=2 F_{1}, \quad C_{2}^2=2 F_{2},\quad  L^2=2 F_3,\quad  M^2=2 F_4,\quad N^2=2 F_5
\]
\[
\begin{array}{l}
F_{1}=-8 (\alpha (A_{2}-A_{1})^2+A_{1} H (A_{2}-B_{1}-2 \delta)+\\
+\delta (H^2-4 \alpha (A_{2}+\beta-\gamma))+A_{1}^2 (B_{1}+\beta-\gamma+\delta) )
\\
F_{2}=8 (\beta (-B_{2}^2+(A_{1}-H)^2-4 \alpha \beta)-A_{2} (B_{2}^2+4 \alpha \beta)+\gamma (B_{2}(B_{2}+H-A_{1})+4 \alpha \beta)-\alpha \gamma^2 )
\\
F_3=8 \beta  A_{1}^2+8 B_{2} F A_{1}-8 F^2 \alpha -8 B_{2}^2 \delta
   -32 \alpha  \beta  \delta
\\
F_4=8 \left(\beta  A_{2}^2-(F (F+\gamma )+2 \beta  (B_{1}+2 \delta ))
   A_{2}+(B_{1}+F) (B_{1} \beta +F (\gamma -\beta ))-(\gamma -2 \beta
   )^2 \delta \right)
\\
F_5=-8 \Big((\beta -\gamma +\delta ) B_{2}^2-H (F+\gamma ) B_{2}+F^2 \alpha
   -H^2 \beta +B_{1} \left(B_{2}^2+4 \alpha  \beta \right)+\\
   +2 F \alpha
   \gamma +\alpha  \left((\gamma -2 \beta )^2+4 \beta  \delta \right)\Big)
\end{array}
\]
The full algebra is given by the following relations:
\[\{ A_{1},\{B_1,B_2\} \}=\{ \{A_1,B_1\},B_{2}\}=8 A_{1} B_{2}-16  \alpha F\]
\[\{\{A_2,B_2\},F\}= \{\{A_{2},F\},B_{2}\}=-8 B_{2} F-16 \beta A_{1}\]
\[ \{ A_{1}, \{A_1,B_1\} \}=\frac{\partial F_{1}}{\partial B_{1}}=-8 A_{1} (A_1-H )+16 \alpha(A_2 - B_{1}) \]
\[ \{\{A_{2},F\},F\}=\frac{\partial F_4}{\partial A_2}=-8 F^2+16 \beta (A_{2}-B_{1})-8 \gamma F-32 \beta \delta\]
\[\{\{A_{1},F\},A_{2} \}=\{\{A_{2},F\},A_{1} \}=8 (B_{1}-A_{2}) B_{2}+8 (A_{1}-H) F\]
\[ \{\{A_{2},F\},A_{2}\}=-\frac{\partial F_4}{\partial F}=8 \gamma (A_{2}-B_{1})+16 A_{2} F+16 (\beta- \gamma) F \]
\[ \{A_{2},\{A_2,B_2\} \}=\frac{\partial F_{2}}{\partial B_{2}}=-16 A_{2} B_{2}-16 (\beta - \gamma ) B_{2}-8 \gamma (A_1-H)\]
\[ \{ \{A_1,B_1\},F\} =\{\{A_{1},F\},B_{1}\}= 8 (B_{1}-A_{2}+2 \delta) B_{2}+8 (F+\gamma) A_{1}\]
\[\{\{A_{1},F\},B_{2}\}=\{\{B_{2},F\},A_{1}\}=\{\{B_1,B_2\},F\}=\{\{A_{2},F\},B_{1}\}=0  \]
\[ \{ B_{1}, \{B_1,B_2\}\}=\frac{\partial F_5}{\partial B_2} =-16 (\beta+\delta-\gamma) B_2+8 (F+\gamma) H-16 (B_1+\beta) B_2\]
\[ \{ \{A_1,B_1\},B_{1} \}=\frac{\partial F_{1}}{\partial A_{1}}=-8 (2 A_1-H) B_1+16 (\gamma-\beta-\delta) A_1+8 (2 \delta H-A_2) H\]
\[ \{\{A_{1},F\},F\}=\frac{\partial F_3}{\partial A_1}=8 B_{2} F+16 \beta A_{1}, \; \{\{B_1,B_2\}, B_{2}\}=\frac{\partial F_5}{\partial B_1}=-8 B_{2}^2-32 \alpha  \beta\]
\[\{ \{A_2,B_2\},B_{1}\}=\{ \{B_1,B_2\},A_{2}\}=8 (A_{2}+B_{1} +2 (\beta -\gamma))B_{2}+ (A_{1}-H) (8 F+8 \gamma )\]
\[ \{\{A_{1},F\},A_{1}\}=-\frac{\partial F_3}{\partial F}=-8 A_{1} B_{2}+16 \alpha F, \; \{ \{A_2,B_2\},B_{2}\}=\frac{\partial F_{2}}{\partial A_{2}}=-8 B_{2}^2-32 \alpha  \beta\]
The second algebra which expand in terms of $C_{1},C_{2},D$  with coefficients any linear combination of integrals $A_{1},A_{2},B_{2},B_{2},F,H$
\[
\{C_{1},C_{2} \}=-8 B_{2} C_{1}+8 (A_{1}-H) C_{2}+8 (H-A_{1}) D
\]
\[
\{C_{1},D \}= -8 B_{2} C_{1} -8 H C_{2}+ 8 (H-A_{1}) D, \; \{C_{2},D \}=8 B_{2} C_{2}-8 B_{2} D
\]
The relation between $B_{1},B_{2},F,A_{2},A_{1},H,M, D$ is:
\[
(B_{2}+F ) C_{1}+ (-A_{2}+B_{1}+H+2 \delta) C_{2}+ (A_{1}+A_{2}-B_{1}-H) D+ (A_{1}+2 \alpha) M =0
\]
\item
\underline{KKM Potential $V_{III}$}
\[
H=p_{x}^2+p_{y}^2+p_{z}^2+ \alpha (x^2+y^2+ z^2)+\frac{\beta}{(x+i y)^2}+\frac{\gamma z}{(x+i y )^3}+\frac{\delta (x^2+y^2-3 z^2)}{(x+i y)^4}
\]
\[
\begin{array}{l}
A_{1}=J^2+\\
+\frac{(x+i y) ( \beta (x+i y) (2 x (x+i y)+z^2)+\gamma z (x^2+y^2+z^2))+\delta (-4 i x y (x+i y)^2-2 z^2 (x^2+y^2)-3 z^4)}{(x+i y)^4}
\end{array}
\]
\[
A_{2}=(J_{2}-i J_{1})^2+\frac{z (\gamma x+i \gamma y-4 \delta z)}{(x+i y)^2}
,\quad
B_{1}=(p_{x}+i p_{y})^2+ \alpha (x+i y)^2-\frac{\delta}{(x+i y)^2}
\]
\[
B_{2}=J_{3} (J_{2}-i J_{1})+\frac{(x+i y)(-4 \beta z (x+i y)+\gamma (2 x (x+i y)-3 z^2))-8 \delta z (x^2+y^2-z^2)  }{4 (x+i y)^3}
\]
\[
F=p_{z} (p_{x}+i p_{y})+\frac{4 \alpha z (x+i y)^4-\gamma (x+i y)+4 \delta z}{4 (x+i y)^3}
\]
The integrals of motion satisfy the following relations:
\[
\{ A_{1},A_{2}\}=\{A_{1},B_{2} \}=\{A_{2},B_{1}\}=\{ B_{1},F \}=0
\]
These relations correspond to the diagram
 \begin{equation}\label{eq:Pi_diagram2}
  \xymatrix{
  A_1 \ar@{--}[d] \ar@{--}[r]
                & A_2 \ar@{--}[d]  \\
  B_2 \ar@{}[dr]
                & B_1 \ar@{--}[d] \\
                & F \ar@{}[u]  }
\end{equation}
If we put
\[
\{A_{1},B_{1}\}=C_{1} ,\; \{ A_{2},B_{2}\}=C_{2} ,\; \{B_{1},B_{2}\}=D,
 \{F,A_{1} \}=L, \; \{ F,A_{2} \}=M%, \; \{ F, B_{2}\}=N
\]
and
\[
C_{1}^2=2F_{1},\; C_{2}^2=2F_{2},\; M^2=2 F_3
\]
then
\[
F_{1}=-8 (\alpha A_{2}^2-\beta H B_{1}+A_{2} (2 \alpha \beta -B_{1} H)+\alpha (\beta-2 \delta)^2-\delta H^2+B_{1}^2 (\delta-\beta)+A_{1} (B_{1}^2+4 \alpha \delta))
\]
\[
F_{2}=2 A_{2}^3+4 \beta A_{2}^2-\frac{\gamma}{2} A_{1}+2 \gamma B_{2} (\beta-2 \delta)+8 \delta B_{2}^2+\frac{1}{2} A_{2} (4 \gamma B_{2}-\gamma^2+4 (\beta-2 \delta)^2+16 \delta A_{1})
\]
\[
F_3=2 A_{2} B_{1}^2-2 F \gamma  B_{1}-\frac{\alpha  \gamma ^2}{2}+8 F^2
   \delta +8 A_{2} \alpha  \delta
\]
The full algebra is given by the following relations:
\[\{\{A_1,B_1\},F\} =\{\{A_{1},F\},B_{1}\}= -8 B_{1} F-4 \alpha \gamma \]
\[ \{ \{A_2,B_2\},B_{1}\}=\{ \{B_1,B_2\},A_{2}\}=16 \delta F-2 \gamma B_{1} \]
\[ \{\{A_{2},F\},B_{1}\}=\{\{B_{2},F\},A_{2}\}=\{ B_{1}, \{B_1,B_2\} \}= 0 \]
\[ \{ A_{1}, \{A_1,B_1\} \}=\frac{\partial F_{1}}{\partial B_{1}}=-16 (A_{1}- \beta + \delta ) B_1+8
  ( A_{2} +8  \beta ) H \]
\[ \{\{A_{1},F\},A_{2} \}=\{\{A_{2},F\},A_{1} \}=2 (4  B_{2}- \beta ) B_{1}+8 (A_{2} + \beta) F \]
\[ \{\{A_2,B_2\},F\}= \{\{A_{2},F\},B_{2}\}=-2 (3 A_{2} + \beta ) B_{1}+2 \gamma F-4 \delta H \]
\[\{ A_{1},\{B_1,B_2\} \}=\{ \{A_1,B_1\},B_{2}\}=8 B_{1} B_{2}-2 \gamma B_{1}+8 (A_{2}+\beta) F\]
\[ \{\{A_{1},F\},B_{2}\}=\{\{B_{2},F\},A_{1}\}=-4 (A_{1}- \beta + \delta) B_{1} +2 (A_{2}+ \beta) H\]
\[ \{\{A_{2},F\},A_{2}\}=-\frac{\partial F_4}{\partial F}=2 \gamma B_{1}-16 \delta F, \; \{\{A_{2},F\},F\}=\frac{\partial F_4}{\partial A_2}= 2 B_{1}^2+8 \alpha \delta \]
\[ \{\{ A_2,B_2\},B_{2}\}=\frac{\partial F_{2}}{\partial A_{2}}=6 A_{2}^2+8 \beta  A_{2}+ 2 \gamma B_{2}+8 \delta A_{1}  +2 \beta ^2+8 \delta^2 -8 \beta \delta -\frac{\gamma ^2}{2}\]
\[ \{\{B_1,B_2\},F\}=-2 B_{1}^2 -8 \alpha \delta, \; \{\{B_{2},F \},B_{1}\}=2 B_{1}^2 +8 \alpha \delta, \; \{\{ B_{2},F\},F\}=2 B_{1} F+\alpha \gamma \]
\[ \{ \{A_1,B_1\},B_{1} \}=\frac{\partial F_{1}}{\partial A_{1}}=-8 B_{1}^2-32 \alpha  \delta, \; \{ A_{2},\{A_2,B_2\} \}=\frac{\partial F_{2}}{\partial B_{2}}=2 \gamma A_{2} + 2 \gamma (\beta  -2 \delta ) +16 B_{2}
   \delta\]
\[ \{\{B_{2},F \},B_{2}\}=(-2  B_{2}+\frac{\gamma}{2}) B_{1}-2 (A_{2}+\beta) F, \; \{\{A_{1},F\},A_{1}\}=2(4 B_{2}- \gamma) H  +16(A_1+\delta- \beta) F \]
\[ \{\{A_{1},F\},F\}=-8 F^2+2 B_{1} H-12 \alpha A_{2}-4 \alpha \beta, \; \{ \{B_1,B_2\}, B_{2}\}=6 A_{2} B_{1}+2 \beta  B_{1}-2 F \gamma +4 H \delta\]
The second algebra which expand in terms of $C_{1},C_{2},D$  with coefficients any linear combination of integrals $A_{1},A_{2},B_{2},B_{2},F,H$
\[\{C_{1},C_{2} \}= 8 B_1 C_2-8 (A_2+\beta) D, \; \{C_{1},D \}=-8 B_{1} D, \; \{C_{2},D \}=-4 \delta C_{1}\]
The relation between $B_{1},B_{2},A_{2},C_{1},C_{2}, D$
\[ B_{1} C_{2}-\frac{8 F \delta  C_{2}}{\gamma }-A_{2} D-D \beta -\frac{C_{1}
   \gamma }{4}+2 D \delta +\frac{4 A_{2} C_{1} \delta }{\gamma }-\frac{8 B_{2} D \delta
   }{\gamma }=0\]

\item \underline{KKM Potential $V_v$}
\[
H=p_{x}^2+p_{y}^2+p_{z}^2+ \alpha (4 x^2+y^2+ z^2)+\beta x+\frac{\gamma}{(y+i z)^2}+\frac{\delta (y-i z)}{(y+i z)^3}
\]
\[
A_{1}=p_{x}^2+4 \alpha x^2+\beta x
,\quad
A_{2}=J_{1}^2+\frac{2 y (\gamma y+i z (\gamma-2 \delta))}{(y+i z)^2}
\]
\[
B_{1}=J_{2} p_{z}-J_{3} p_{y}+\frac{\beta}{4} (y^2+z^2)+x (\alpha (y^2+z^2)-\frac{2 \delta y}{(y+i z)^3}+\frac{\delta-\gamma}{(y+i z)^2})
\]
\[
B_{2}=(p_{z}-i p_{y})^2+\frac{\delta -\alpha (y+i z)^4}{(y+i z)^2}
\]
\[
F=(p_{z}- i p_{y}) (J_{2}+i J_{3})-\frac{1}{4}(y+i z)^2 (4 \alpha x+\beta)-\frac{\delta x}{(y+i z)^2}
\]
The indergrals satisfy the equation:
\[
\{ A_{1},A_{2}\}=\{A_{1},B_{2} \}=\{A_{2},B_{1}\}=\{A_{2},F\}=\{B_{1},F\}=0
\]
The corresponding diagram is the same as in the case of the potential KKM $V_{III}$ see (\ref{eq:Pi_diagram2}).
\[
\{A_{1},B_{1}\}=C_{1} ,\; \{ A_{2},B_{2}\}=C_{2} ,\; \{B_{1},B_{2}\}=D,
\;
 \{F,A_{1} \}=L, \; \{ F,A_{2} \}=M, \; \{ F, B_{2}\}=N
\]
\[
C_{1}^2=2F_{1},\quad C_{2}^2=2F_{2},\quad M^2= 2 F_4
\]
\[
F_{1}=8 ( \gamma (B_{1}(A_{2}+B_{1}-H)-\alpha \gamma)+\delta (-B_{1}^2+(A_{2}-H^2)+4 \alpha \gamma)-4 \alpha \delta^2-A_{1} (B_{1}^2+4 \alpha \delta))
\]
\[
F_{2}=\frac{1}{2} (4 A_{2}^3-8 A_{2}^2 H-16 \alpha B_{2}^2+4 \beta H B_{2} -\beta^2 (A_{1}-\gamma+\delta)+4 A_{2} (H^2-\beta B_{2}-4 \alpha (A_{1}-\gamma+\delta)))
\]
\[
F_3=2 A_{2} B_{1}^2+2 F \beta  B_{1}-8 F^2 \alpha +\frac{\beta ^2 \delta
   }{2}+8 A_{2} \alpha  \delta
\]
The full algebra is given by the following relations:
\[ \{ \{A_2,B_2\},B_{2}\}=\frac{\partial F_{2}}{\partial A_{2}}=-8 B_{2}^2-32 \alpha  \delta  \]
\[\{\{A_2,B_2\},F\}= \{\{A_{2},F\},B_{2}\}=-8 B_{2} F-4 \beta \delta \]
\[\{ A_{1},\{A_1,B_1\} \}=\frac{\partial F_{1}}{\partial B_{1}}=-16\alpha  B_{1} -2 \beta (A_{1} - H )\]
\[ \{ A_{1},\{B_1,B_2\} \}=\{\{A_1,B_1\},B_{2}\}=-16 \alpha F+2 \beta B_{2}\]
\[ \{ \{ B_{1}, F\},A_{1} \}=\{\{A_{1},F\},B_{2}\}=\{\{B_1,B_2\}, B_{2}\}=0\]
\[ \{\{A_{1},F\},A_{2}\}=\{\{A_{2},F\},A_{1} \}=8 B_{1} B_{2}+8 (A_{1}-H) F \]
\[ \{\{A_2,B_2\},B_{1}\}=\{ \{B_1,B_2\},A_{2}\}=8 B_{1} B_{2} +8 (A_{1}-H) F\]
\[ \{A_{2},\{A_2,B_2\}\}=\frac{\partial F_{2}}{\partial B_{2}}=-16 (A_2 +\delta -\gamma) B_2+ 8 \gamma (A_1-H)\]
\[ \{\{A_1,B_1\},F\} =\{\{A_{1},F\},B_{1}\}= -2 (3 A_1-H) B_2-2 \beta F+4 \alpha \gamma \]
\[  \{\{A_{2},F\},B_{1}\}=\{\{B_{1},F\},A_{2}\}=-4 A_{2} B_{2}+2 \gamma A_{1}+4 (\gamma -\delta)B_{2}-2 \gamma H \]
\[ \{ A_1,B_1\},B_{1}\}=\frac{\partial F_{1}}{\partial A_{1}}=2 H^2+2 (3 A_1-4 H) A_1-8 \alpha  A_{2} -2 \beta B_{1}
+8 \alpha  (\gamma -  \delta) \]
\[  \{\{A_{2},F\},A_{2}\}=16 A_2 F+8 \gamma B_{1}+16 (\delta -\gamma) F , \quad \{\{ B_{1},F\},F\}=2 B_{2} F+\beta \delta \]
\[ \{\{A_{1},F\},A_{1}\}=-\frac{\partial F_3}{\partial F}= 16 \alpha F-2 \beta B_{2}, \;  \{\{A_{1},F\},F\}=\frac{\partial F_3}{\partial A_1} = 2 B_{2}^2+8 \alpha \delta \]
\[  \{\{B_1,B_2\},F\}=\{\{B_{1},F\},B_{2}\}= 2 B_{2}^2+8 \alpha \delta ,\quad \{\{B_{1},F \},B_{1}\}=-2 B_{1} B_{2}-2 A_{1} F+2 F H\]
\[ \{\{A_{2},F\},F\}=-8 F^2+2 \gamma B_{2}+12 \delta A_{1}-4 \delta H , \; \{ B_{1}, \{B_1,B_2\} \}= 2 (3 A_{1} - H )B_{2}+2 \beta  F -4 \alpha  \gamma\]
The second algebra which expand in terms of $C_{1},C_{2},D$  with coefficients any linear combination of integrals $A_{1},A_{2},B_{2},B_{2},F,H$
\[
\{C_{1},C_{2} \}=-8 B_{2} C_{1}+8 (H-A_{1}) D, \quad \{C_{1},D \}=-4 \alpha C_{2}, \quad \{ C_{2},D \}=-8 B_{2} D
\]
The relation between $A_{1},B_{1},B_{2},C_{1},C_{2},D,H,F$
\[
B_{2} C_{1}-\frac{8 F \alpha  C_{1}}{\beta }+A_{1} D-D H+\frac{C_{2} \beta
   }{4}+\frac{4 A_{1} C_{2} \alpha }{\beta }+\frac{8 B_{1} D \alpha }{\beta }=0
\]

\item \underline{KKM Potential $V_{iv}$}
\[
H=p_{x}^2+p_{y}^2+p_{z}^2+ \alpha (4 x^2+y^2+ z^2)+\beta x+\frac{\gamma}{y^2}+\frac{\delta}{z^2}
\]
\[
A_{1}=p_{x}^2+4 \alpha x^2+\beta x
,\;
A_{2}=p_{y}^2+\alpha y^2+\frac{\gamma}{y^2}
,\;
B_{1}=J_{2} p_{z}+\alpha x z^2+\frac{\beta z^2}{4}-\frac{\delta x}{z^2}
\]
\[
B_{2}=J_{1}^2+\frac{\gamma z^2}{y^2}+\frac{\delta y^2}{z^2}
,\;
F=p_{y} J_{3}-\alpha x y^2-\frac{\beta y^2}{4}+\frac{\gamma x}{y^2}
\]
The above integrals satisfy the relations:
\[
\{ A_{1},A_{2}\}=\{A_{1},B_{2} \}=\{A_{2},B_{1}\}=0
\]
These relations correspond to the diagram:
 \begin{equation}\label{eq:Pi_diagram3}
  \xymatrix{
  A_1 \ar@{--}[d] \ar@{--}[r]
                & A_2 \ar@{--}[d]  \\
  B_2 \ar@{}[dr]
                & B_1  \\
                & F \ar@{}[u]  }
\end{equation}

 We can define
\[
\{A_{1},B_{1}\}=C_{1} ,\qquad \{ A_{2},B_{2}\}=C_{2}
\]
and
\[
C_{1}^2=2F_{1},\quad C_{2}^2=2F_{2}
\]
where
\[
F_{1}=-8 \alpha B_{1}^2+2 (A_{1}+A_{2}-H)(A_{1}(A_{1}+A_{2}-H)-\beta B_{1})-\frac{\delta}{2} (16 \alpha A_{1}+\beta^2)
\]
\[
F_{2}=-8 ( \alpha B_{2}+\gamma A_{1}^2-2 A_{1} H (B_{2}+2 \gamma)+A_{1} (-2 \gamma H+A_{2}(B_{2}+2 \gamma))+A_{2}^2 (B_{2}+\gamma+\delta)+\gamma (H^2-4 \alpha \delta) )
\]
The full parafermionic-like algebra is given by the following relations:
\[ \{ \{ B_{1}, F\},A_{1} \}=0 \]
\[ \{\{B_1,B_2\},B_{2}\}=-8 (B_{1} +F )B_{2}-16 B_{1} \gamma -16 F \delta \]
\[ \{\{A_{1},F\},A_{2} \}=\{\{A_{2},F\},A_{1} \}=-16 \alpha F-2 \beta A_{2}\]
\[ \{ B_{1},\{B_1,B_2\} \}= 2 (3 A_1+A_2-H) B_2-8 B_{1}F+4  \delta A_{2}\]
\[ \{\{B_{2},F \},B_{1}\}=2 (3 A_{1}+ B_2+2 \delta )A_{2} -8 B_{1} F-2 B_{2} H\]
\[ \{ A_{1}, \{A_1,B_1\} \}=\frac{\partial F_{1}}{\partial B_{1}}=-16 \alpha  B_{1} -2 \beta (A_{1}  +A_{2}  -H )  \]
\[ \{ A_{2},\{A_2,B_2\} \}=\frac{\partial F_{2}}{\partial B_{2}}=-8 (A_1+A_{2}-H ) A_{2}-16 \alpha B_{2} \]
\[ \{\{B_1,B_2\},F\}=2 (A_{2}-2 A_{1} )B_{2}+8 B_{1} F+4 \gamma (A_{1}+A_{2}-H)\]
\[ \{\{A_1,B_1\},F\} =\{\{A_{1},F\},B_{1}\}= 2 (A_{1}+ A_{2}- H) A_2+4 \alpha B_{2}\]
\[ \{\{A_{1},F\},B_{2}\}=\{\{B_{2},F\},A_{1}\}=8 (A_{1}+A_{2}-H) F -8 A_{2}B_{1}\]
\[ \{ \{A_2,B_2\},B_{1}\}=\{ \{B_1,B_2\},A_{2}\}=8 (A_{1}+A_{2}-H) F-2 \beta B_{2}\]
\[ \{ A_{1},\{B_1,B_2\} \}=\{ \{A_1,B_1\},B_{2}\}=8 (A_{1}+A_{2}-H) F-8 A_{2} B_{1} \]
\[ \{\{A_{2},F\},B_{1}\}=\{\{B_{1},F\},A_{2}\}=-2 (A_1+A_2-H) A_2-4 \alpha B_{2} \]
\[ \{\{B_{1},F\},B_{2}\}=2 (A_1+2 A_2-H) B_2+4 \gamma (A_{1}+A_{2}-H)+4 \delta A_{2}\]
\[ \{ \{A_2,B_2\},B_{2}\}=\frac{\partial F_{2}}{\partial A_{2}}=-8 (A_1+2 A_2-H) B_2-16 \gamma (A_1+A_2-H)-16 \delta A_{2} \]
\[ \{ \{A_1,B_1\},B_{1} \}=\frac{\partial F_{1}}{\partial A_{1}}=2 (3 A_1+4 A_2-4 H) A_1+2 (A_2-2 H) A_2+2 H^2-2\beta B_{1}  -8 \alpha  \delta\]
\[ \{\{B_{2},F\},A_{2}\}=8 (H-A_{1}-A_{2}) F+2 \beta B_{2}, \; \{\{A_{2},F\},B_{2}\}=8 (H-A_{1}-A_{2}) F+8 A_{2} B_{1}\]
\[ \{\{B_{1},F \},B_{1}\}=2 (H-A_{1}-A_{2}) F+\frac{\beta}{2} B_{2}, \; \{\{B_{2},F \},B_{2}\}=8 (B_{1}+ F) B_2+16 \gamma B_{1}+16 \delta F\]
\[ \{\{A_{1},F\},F\}=2 (A_{2}-2 A_{1} )A_{2}+2 \beta F-8 \alpha \gamma, \; \{\{A_{2},F\},F\}=-2 (A_{2}- A_{1}) A_{2}-2 \beta F+8 \alpha \gamma \]
\[ \{\{ B_{1},F\},F\}=-2 A_{2} B_{1}+\frac{\beta}{2} B_{2}, \; \{\{ B_{2},F\},F\}=2 (2 A_{1} - A_{2} ) B_{2}+4 \gamma (H-A_{1}-A_{2})-8 B_{1} F\]
\[  \{\{A_{1},F\},A_{1}\}=16 \alpha F+2 \beta A_{2}, \; \{\{A_{2},F\},A_{2}\}=16 \alpha F+2 \beta A_{2},\;\{\{A_2,B_2\},F\}= 8 A_{2} B_{1}-2 \beta B_{2}\]
The second algebra which expand in terms of $C_{1},C_{2},D$  with coefficients any linear combination of integrals $A_{1},A_{2},B_{2},B_{2},F,H$
\[
\{ C_{1},C_{2}\}= -8 A_{2} C_{1}+2 \beta C_{2} +16 \alpha D
\]
\[
\{ C_{1},D \}=-8 F C_{1}+ 2 (3 A_{1}+A_{2}-H) C_{2}-2 \beta D, \; \{ C_{2},D \}= 8(B_{2}+2 \gamma) C_{1}+8 F C_{2} -8 A_{2} D
\]
The relation between $C_{1}, C_{2}, A_{1},B_{1},B_{2}, F, M$
\[
\begin{array}{c}
\displaystyle \frac{1}{4} (4 B_{2}+4 F+8 \gamma )  C_{1}+ \frac{1}{4} (-2 A_{1}-4 B_{1}+4 F) C_{2}+ \\ \displaystyle + \frac{1}{4} (4 A_{1}-4 H+\beta ) D+ \frac{1}{4} (-4 B_{1}+4 B_{2}+8 \delta ) M=0
\end{array}
\]

\item \underline{KKM Potential $V_{vi}$}
\[
H=p_{x}^2+p_{y}^2+p_{z}^2+\alpha (-2 (x-i y)^3+4 \alpha (x^2+y^2)+z^2)+\beta (-3 (x-i y)^2+2 (x+i y))+\gamma (x-i y)+\frac{\delta}{z^2}
\]
\[
A_{1}=(p_{x}- i p_{y})^2+ 4 (x- i y )(\alpha (x-i y)+\beta)
,\;
A_{2}=p_{z}^2+\alpha z^2+\frac{\delta}{z^2}
\]
\[
\begin{array}{l}
\displaystyle B_{1}=J_{3} (p_{x}-i p_{y})-\\
-\frac{i}{4} \Bigl ( (p_{x}+i p_{y})^2+\frac{i}{4} ( 3 \alpha x^4+4 x^3 (\alpha+\beta-3 \alpha i y)+4 x y (\alpha(3 i y^2+y-2 i))-3 \beta y)+\\
 + 2 \gamma i x (i+y)- x^2(2 \alpha (y(2 i+9 y)-2)+4 \beta (3 i y-2)+\gamma )+y (y-2 i) (3 \alpha y^2+2 i y (\alpha+2 \beta)+\gamma ) \Bigr )
\end{array}
\]
\[
B_{2}= ( J_{2}+i J_{1}) p_{z}+ z^2 (\alpha (x- i y)+\beta)-\frac{\delta (x-i y)}{z^2}
\]
\[
F=(J_{2}+i J_{1})^2+z^2 ( 3 \alpha x^2-3\alpha y^2+2 \beta-4 i y (\alpha+\beta)+x(4 \beta-6 \alpha i y)-\gamma)+\delta ((x-i y)^2+4 i y ) )
\]
The integrals satisfy the relations
\[
\{ A_{1},A_{2}\}=\{A_{1},B_{2} \}=\{A_{2},B_{1}\}=0
\]
The "$\Pi$ shape diagram corresponding to the above relations is given by (\ref{eq:Pi_diagram3}).
\[
\{A_{1},B_{1}\}=C_{1} ,\quad \{ A_{2},B_{2}\}=C_{2}
\]
\[
C_{1}^2=2F_{1},\quad C_{2}^2=2F_{2}
\]
\[
F_{1}=-2 (A_{1}^3-2 \gamma A_{1}^2+A_{1}(16 \alpha i B_{1}-4 \beta A_{2}+4 \beta H+\gamma^2)-4 (-4 \beta^2 i B_{1})+(A_{2}-H)(\alpha A_{2}-\alpha H-\beta \gamma) )
\]
\[
F_{2}=-8 \alpha B_{2}^2+2 A_{2} (A_{1} A_{2}+4 \beta B_{2})-8 \delta (\alpha A_{1}+\beta^2)
\]

\[ \{\{A_{1},F\},A_{1}\}=0\]
\[  \{ A_{2},\{A_2,B_2\}\}=\frac{\partial F_{2}}{\partial B_{2}}=8 A_{2} \beta -16 B_{2} \alpha\]
\[  \{\{B_{2},F\},A_{2}\}=-8 ( A_{1} +2 \beta -\gamma )B_{2}+8 \beta F\]
\[ \{\{A_{2},F\},A_{2}\}=-8 (A_{1} +2 \beta-\gamma) A_{2}+16 \alpha F \]
\[ \{\{A_{1},F\},A_{2} \}=\{\{A_{2},F\},A_{1} \}=64 \alpha B_{2}-32 \beta A_{2}\]
\[ \{ B_{1}, \{B_1,B_2\}\}=-2 A_{1} B_{2}+2 (\gamma -2 \beta) B_2+2 \beta F \]
\[  \{\{A_{1},F\},B_{1}\}= i(8 A_{1} A_{2}+32 \alpha B_{2}-16 \alpha F-8 \gamma A_{2})\]
\[ \{\{A_{2},F\},B_{2}\}=8 (2 A_{2}+ A_{1} - H ) A_2+8 \beta F-32 \alpha \delta \]
\[\{ A_{1},\{B_1,B_2\}\}=\{\{A_1,B_1\},B_{2}\}= i (16 \alpha B_{2}-8 \beta A_{2})\]
\[ \{ \{ B_{1}, F\},A_{1} \}=i(-16 A_{1} A_{2}-32 (\alpha +\beta) B_{2}+16 \beta A_{2})\]
\[ \{\{B_{2},F \},B_{2}\}=8 (A_{1} + 2 A_{2}- H) B_2-4 A_{1} F-16 \beta \delta \]
\[\{\{B_1,B_2\},B_{2}\}=-2 i (2 A_2-H) A_2+ 2 \beta i (2 B_2-F) +8i\alpha  \delta\]
\[
      \begin{array}{cl}
       \{\{ B_{1},F\},F\}=& -8 (2 i A_2-4 B_1+10 i B_2-2 i F-H) A_2-\\
                          &- 8 (4 i B_2+i A_1- 8 B_1- 4 i H) B_2 \\
                          & 4 i (A_1-2 H) F+8 i (2 \beta-\gamma)+4 i (\gamma-2 \beta) F-\\
                          &-48 i \delta A_1+16 i \delta (2\alpha+4 \beta+\gamma)
   \end{array}
\]
\[ \{\{A_2,B_2\},F\}= 8 (2 A_{2}+ A_{1}- H) A_2 +8 (A_{1} + 2\beta -\gamma ) B_{2}-32 \alpha \delta \]
\[ \{\{A_1,B_1\},F\} = i(24 A_{1} A_{2} +32 (\alpha +\beta) B_{2}-8 (2 \beta +\gamma) A_{2} -16 \alpha F)\]
\[ \{\{A_2,B_2\},B_{1}\}=\{\{B_1,B_2\},A_{2}\}=i (2 A_{1} A_{2}+8 \alpha B_{2}-4 \alpha F-2 \gamma A_{2})\]
\[ \{\{A_{1},F\},B_{2}\}=\{\{B_{2},F\},A_{1}\}=i(8 A_{1} A_{2}+32 \alpha B_{2}-16 \alpha F-8 \gamma A_{2})\]
\[
\begin{array}{cl}
\{\{B_1,B_2\},F\}=& -4 (2 i A_2-4 B_1+2 i B_2- i H) A_2+2 i (2 B_2-F) A_1+\\ &+4 i(2\beta -\gamma) B_{2}+2 i(\gamma -2 \beta) F+16 i \alpha \delta+16 i \beta \delta
\end{array}
\]
\[ \{\{A_1,B_1\},B_{1} \}=\frac{\partial F_{1}}{\partial A_{1}}=-6 A_{1}^2+8 \gamma  A_{1}-32 \alpha i B_{1} +8
  \beta A_{2}  -8\beta H -2 \gamma ^2\]
\[ \{\{A_{1},F\},F\}=32 (2 A_2+A_1-H) A_2+32 A_{1}B_{2}+  32 (2 \beta-\gamma) B_2 -128 \alpha  \delta \]
\[
        \begin{array}{cl}
 \{\{A_{2},F\},F\}= &16 (4 A_2+A_1+4 i B_1+2 B_2-2 H) A_2+ (A_1+2 \beta-\gamma) F+\\ & -64 \delta (2 \alpha+\beta)
      \end{array}
      \]
\[
       \begin{array}{cl}
       \{\{ B_{2},F\},F\}= & 16 (2 B_2+A_1+4 A_2+4 i B_1-H) B_2 -8 (A_{1} +2 A_{2} -
   H ) F +\\ & +48 \delta A_{1}  -16 \delta ( 2 \beta + \gamma )
       \end{array}
\]
\[
\begin{array}{cl}
\{\{B_{1},F\},B_{2}\}=& -4 i (2 A_2+4 B_2-H) A_2-4 i (2 B_2-F) A_1+8 i (H+\beta) B_2-\\ &-4 \beta i F+16 i \delta (\alpha+\beta)
\end{array}
\]
\[\{\{B_{2},F \},B_{1}\}=-8 (2 B_1+i B_2) A_2+6 i (F-2 B_2) A_1+4 i (2 H+\gamma) B_2-2 \gamma i F\]
\[ \{\{B_{1},F \},B_{1}\}=-8 (A_1+A_2-H) B_2+ 6 A_1 F+16 i A_2 B_1+8 \beta B_2-2 (\gamma+2 \beta) F \]
\[ \{\{A_2,B_2\},B_{2}\}=\frac{\partial F_{2}}{\partial A_{2}}=4 A_{1} A_{2}+8 B_{2} \beta, \;   \{ A_{1}, \{A_1,B_1\}\}=\frac{\partial F_{1}}{\partial B_{1}}=-32 i \beta^2-32 i A_{1} \alpha \]
\[ \{\{A_{2},F\},B_{1}\}=\{\{B_{1},F\},A_{2}\}=i(4 A_{1} A_{2}+8 A_{1} B_{2}+8 (2 \alpha+2\beta -\gamma) B_{2}-8 (\alpha+\beta) F-4 \gamma A_{2}) \]
The second algebra which expand in terms of $C_{1},C_{2},D$  with coefficients any linear combination of integrals $A_{1},A_{2},B_{2},B_{2},F,H$
\[
\{ C_{1},C_{2} \}=16 \alpha i C_{2},  \quad \{ C_{1},D \}=8 \beta C_{2}-16 \alpha i D, \quad \{ C_{2},D \}=-2 A_{2} C_{1}
\]
There exist an relation between $A_{1}, A_{2},C_{1},C_{2},B_{2}, D, H$
\[
- i \frac{1}{\beta \gamma} (\beta A_2-2 \alpha B_2) C_1-\frac{1}{\beta \gamma}  (2 \alpha A_2+\beta A_1- 2\alpha H- \beta \gamma) C_2+ i \frac{4}{\beta \gamma} (\alpha A_1+\beta^2) D=0
\]

\item \underline{KKM potential $V_{vii}$}
This potential is somehow exceptional because is the only one, where the integrals don't satisfy a "$\Pi$" shape diagram.
\[
\begin{array}{rl}
 H=& \displaystyle p_{x}^2+p_{y}^2+p_{z}^2+(x+i y) \alpha + (\frac{3}{4} (x+i y)^2+\frac{z}{4} ) \beta +\\ &  \displaystyle
   +((x+i y)^3+\frac{3}{4} z (x+i y)+ \frac{1}{16} (x-i y))
   \gamma+ \\ & + \displaystyle (\frac{5}{16} (x+i y)^4+\frac{3}{8} z (x+i
   y)^2+\frac{1}{16} (x^2+y^2+z^2)) \delta
\end{array}
\]
The integrals of motion are given by:
\[
 A_{1}=(p_{x}+i p_{y})^2+\frac{1}{16} (x+i y) (2 \gamma +\delta (x+i y))
\]
\[
A_{2}=p_{z} (p_{x}+i p_{y})+\frac{1}{16} (2 \beta i y-3 \gamma y^2+\gamma z+\delta x^3- \delta i y^3+\delta i y z +3 x^2 \gamma(1+i \delta)+x (2 \beta+6 \gamma i y-3 \delta y^2+\delta z))
\]
\[
\begin{array}{l}
 B_{1}=J_{2} p_{z}-J_{3} p_{y}+i (J_{3} p_{x}-J_{1} p_{z})-\frac{i}{2} py pz
 +  \frac{1}{64} (-8 \delta  x^5-8 (3 \gamma +5 i y \delta ) x^4- \\  -4
   \left(-20 \delta  y^2+24 i \gamma  y+4 \beta +z \delta \right)
   x^3 +\left(80 i \delta  y^3+144 \gamma  y^2-2 i (24 \beta +6 z \delta
   +\delta ) y-16 \alpha +\gamma \right) x^2+ \\ +\left(-40 \delta  y^4+96 i
   \gamma  y^3+4 (12 \beta +3 z \delta +\delta ) y^2-2 i (16 \alpha +3
   \gamma ) y+4 z (2 \beta +z \delta )\right) x-24 y^4 \gamma +\\   +y^2 (16
   \alpha +7 \gamma )+z (16 \alpha +(8 z-1) \gamma )-8 i y^5 \delta +2 i
   y (2 z-1) (2 \beta +z \delta )+2 i y^3 (8 \beta +2 z \delta +\delta
   ))
\end{array}
\]
\[
\begin{array}{ll}
\displaystyle B_{2}=&  i \Bigl (J_{3} p_{z}+i J_{1} p_{y}+i J_{2} p_{x}+\frac{i}{4} pz^2-\frac{1}{64} i (x^2+2 i y x-y^2-z ) \\ &   (3 \delta  x^2+8
   \gamma  x+6 i y \delta  x+4 \beta +8 i y \gamma -3 y^2 \delta +z
   \delta ) \Bigr )
\end{array}
\]
\[
\begin{array}{l}
 F= \frac{1}{4} \left(4 J_{1}^2+8 i (J_{2}-p_{x}) J_{1}-4
   J_{2}^2+p_{y}^2+4 J_{2} (p_{x}-i
   p_{y})+p_{z} (p_{z}-12 i J_{3})\right)+\\ +\frac{1}{64} ((16 z-3) \delta  x^4   + 8 (6 z-1) \gamma  x^3+2 (2 (8
   z-1) \beta +z (6 z+1) \delta ) x^2+\\+
   2 z (16 \alpha +(8 z+3) \gamma )
   x+y^4 (16 z+5) \delta +z (4 \beta +z \delta )- \\  -4 i y^3 (4 (3 z \gamma
   +\gamma )+x (16 z+3) \delta )-y^2 (6 (16 z+1) \delta  x^2+\\+24 (6
   z+1) \gamma  x+4 (8 z+3) \beta +(6 z (2 z+1)-1) \delta )+ \\  +i y
   (4 (16 z-1) \delta  x^3+144 z \gamma  x^2+8 (8 z \beta +\beta
   +z (3 z+1) \delta ) x+\\+16 (2 z \alpha +\alpha )+(2 z (8 z+7)-1)
   \gamma ))
\end{array}
\]
This system satisfy the relations
\[
\{ A_{1},A_{2}\}=\{A_{1},B_{2} \}= \{ B_{1}, F \}=0
\]
and the corresponding diagram is
 \begin{equation}\label{eq:Pi_diagram4}
  \xymatrix{
  A_1 \ar@{--}[d] \ar@{--}[r]
                & A_2   \\
  B_2 \ar@{}[dr]
                & B_1 \ar@{--}[d] \\
                & F \ar@{}[u]  }
\end{equation}
The next relations correspond to the above diagram
\[
 \{ A_{2},B_{2}\}=C_{2} ,\; C_{2}^2=2 F_{2}
\]
\[
 F_{2}=  \frac{1}{2}A_{1}^3+\frac{\beta}{16} A_{1}^2+\frac{\beta^2}{512} A_{1}-\frac{\gamma}{32} A_{1} A_{2}-\frac{\beta \gamma}{512} A_{2}-\frac{\gamma^2}{512} B_{2}-\frac{\delta}{32} A_{1} B_{2}-\frac{\delta}{128} A_{2}^2
\]
The full algebra is given by the following relations:
\[ \{ \{B_{1}, A_{2} \},B_{2} \}=\frac{1}{8} (16 A_1+\beta) A_1-\frac{\gamma}{16} A_2 \]
\[ \{A_{1}, \{A_1,B_2\} \}=\{A_{1},\{A_1,B_1\}\}=\{ \{A_{1}, F \}, B_{2}\}=0  \]
\[ \{ \{F,B_{2}\}, A_{2} \}=\frac{1}{8} (16 A_1-\beta) A_2+\frac{\alpha}{2} A_1-\frac{3 \gamma}{8} B_2-\frac{\gamma}{32} H \]
\[   \{ \{A_{1}, F \}, B_{2} \}= \{ \{B_{2}, F \}, A_{1} \}=\frac{1}{4} (16 A_1+\beta) A_1-\frac{\gamma}{8} A_2 \]
\[ \{\{F,A_{1}\}, A_{2} \}= \{\{F,A_{2}\}, A_{1} \}=-\frac{\gamma}{128} (16 A_1+\beta)-\frac{\delta}{16} A_2\]
\[ \{\{B_1,B_2\}, A_{1} \}=\{ B_{2},\{A_1,B_1\}\}=-\frac{ \gamma }{16}A_{1}-\frac{
   \delta }{32}A_{2}-\frac{\beta  \gamma }{256}\]
\[
\{ \{F,B_{2}\},B_{2}\}=-4 A_2^2-\frac{1}{64} (16 \alpha -\gamma) A_2-8 A_1 B_2+\frac{\gamma}{16} B_1
\]
\[
\{ \{F,A_{1}\},A_{1}\}= -\frac{ \delta }{8} A_1-\frac{\gamma ^2}{128},  \{\{F,B_{1}\}, A_{1} \}=\frac{ \gamma }{8} A_1+\frac{
   \delta }{16} A_2+\frac{\beta  \gamma }{128}\;
\]
\[   \{\{F,A_{1}\}, B_{1} \}=  \frac{1}{32} (384 A_1+8 \beta+\delta) A_2+\frac{1}{32} (48 \alpha+\gamma) A_1-\frac{\gamma}{4} B_2+\frac{\delta}{16} B_1-\frac{\gamma}{16} H+\frac{\beta \gamma}{512}+\frac{\alpha \beta}{32}\]
\[\{\{A_2,B_2\},B_{2}\}=\frac{\partial F_{2}}{\partial A_{2}}=-\frac{\gamma }{32} A_1-\frac{ \delta }{64} A_2-\frac{\beta  \gamma
   }{512}, \; \{A_{2},\{A_2,B_2\}\}=\frac{\partial F_{2}}{\partial B_{2}}=-\frac{\delta }{32} A_1 -\frac{\gamma ^2}{512}\]
\[ \{ \{B_{1},A_{2}\},A_{2}\}=\frac{\gamma }{16} A_1+\frac{\delta }{32} A_2+\frac{\beta  \gamma
   }{256}, \; \{ \{ B_{1}, A_{2} \},A_{1} \}=\{ A_{2},\{A_1,B_1\}\}=\frac{ \delta }{16} A_{1} +\frac{\gamma ^2}{256}\]
\[\{\{A_2,B_2\},B_{1}\}=-\frac{1}{128} (384 A_1+32 \beta+\delta) A_1+\frac{3\delta }{32} B_2+\frac{ \gamma }{8} A_2+\frac{\delta }{128} H -\frac{1}{2048} (8\beta^2+\gamma^2)+\frac{\alpha  \gamma}{128}\]
   \[\{\{B_1,B_2\}, A_{2} \}=-\frac{1}{128}(128 A_1+16 \beta+\delta) A_1 +\frac{3  \delta }{32}B_{2}+\frac{ \gamma }{16}A_{2}+\frac{\delta }{128}H -\frac{1}{2048} (8 \beta^2+\gamma^2) +\frac{\alpha  \gamma
   }{128} \]
\[  \{\{A_1,B_1\},F \}= -\frac{1}{32} (384 A_1+8 \beta+\delta) A_2-\frac{1}{32} (48 \alpha+\gamma) A_1+\frac{\gamma}{4} B_2-\frac{\delta}{16} B_1+\frac{\gamma}{16} H-\frac{\beta \gamma}{512}-\frac{\alpha \beta}{32}\]
\[
       \begin{array}{cl}
 \{ F, \{B_{1},A_{2}\}\}=& \displaystyle -\frac{1}{128} (384 A_1+512 B_2+128 H+\delta-24 \beta) A_1 -\frac{1}{128} (8 F+H)-\\ & \displaystyle -\frac{1}{32} (8 \beta+96 \delta) B_2+\frac{\beta}{16} H-\frac{1}{32} (256 A_2+48 \alpha+96 \gamma) A_2- \\ & \displaystyle -\frac{1}{4096} (\gamma^2-256 \alpha^2)
\end{array}
   \]
\[
\begin{array}{cl}
\{ \{F,A_{2}\},B_{1}\}= & \displaystyle -\frac{1}{128} (A_1+512 B_2+128 H+\delta +8 \beta) A_1+ \frac{1}{32} (256 A_2+48 \alpha +\gamma) A_2 \\
& \displaystyle -\frac{1}{32} (8 \beta +\delta) B_2-\frac{1}{128} (8 \beta -\delta) H -\frac{\delta}{16} F
\end{array}
\]
\[ \{F,\{B_{2},A_{2}\}\}=  \frac{1}{64} (384 A_1+8 \beta+\delta) A_2+\frac{1}{64} (48 \alpha+\gamma) A_1-\frac{\gamma}{8} B_2+\frac{\delta}{32} B_1-\frac{\gamma}{32} H+\frac{\beta \gamma}{1024}+\frac{\alpha \beta}{64} \]
\[ \{\{F,A_{2}\},B_{2}\}=  -\frac{1}{64} (256 A_1+16 \beta+\delta) A_2-\frac{1}{64} (16 \alpha+\gamma) A_1-\frac{\gamma}{4} B_2-\frac{\delta}{32} B_1-\frac{\beta \gamma}{1024}-\frac{\alpha \beta}{64} \]
\[
\begin{array}{cl}
\{ F, \{B_{1},B_{2}\}\}= & -\{ \{ F,B_{2}\},B_{1}\}=  \displaystyle -\frac{1}{128} (256 B_1-16 \alpha+\gamma) A_1-\frac{\gamma}{8} F-\frac{\beta}{8} B_1- \\ & \displaystyle -\frac{1}{128} (128 A_2-\gamma+12 \alpha) H-\frac{1}{32} (384 A_2+48 \alpha +5 \gamma) B_2
\end{array}
\]
\[
\begin{array}{cl}
        \{ F,\{F,B_{2}\}\}= & \displaystyle -\frac{1}{32} (32 A_1+320 B_2+384 F-48 H+\beta) A_1-\frac{\beta}{4} F-\frac{1}{32} (16 \alpha+\gamma) B_1 \\ & \displaystyle -\frac{1}{8} (192 B_2+64 H+\beta) B_2-\frac{1}{2} H^2+\frac{\beta}{32} H
        \end{array}
\]
\[
\begin{array}{cl}
        \{ F,\{F,A_{1}\}\}= & \displaystyle -\frac{1}{64} (128 A_1+512 B_2+128 H+8 \beta+\delta) A_1+\frac{1}{64} (\delta-8 \beta) H-\frac{\delta}{8} F\\ & \displaystyle +\frac{1}{16} (256 A_2+48 \alpha+\gamma) A_2-\frac{1}{16} (8 \beta+\delta) B_2+\frac{1}{2048} (256 \alpha^2-\gamma^2)
        \end{array}
\]
\[ \{\{A_1,B_1\},B_{1}\}=\frac{1}{64} (384 A_1+32 \beta+\delta) A_1-\frac{\gamma}{4} A_2-\frac{3 \delta}{16} B_2-\frac{\delta}{64} H+\frac{1}{1024} (8 \beta^2+\gamma^2)-\frac{\alpha \gamma}{64}\]
\[ \{B_{1},\{B_{1},A_{2}\}\}=\frac{1}{64} (384 A_1+8 \beta+\delta) A_2+\frac{1}{64} (\gamma +48 \alpha) A_1-\frac{\gamma}{8} B_2-\frac{\gamma}{32} H+\frac{\delta}{32} B_1+\frac{\alpha \beta}{64}+\frac{\beta \gamma}{1024}\]
\[
 \begin{array}{cl}
\{B_{1},\{B_1,B_2\}\}=& \displaystyle -\frac{1}{32} (32 A_1-64 B_2-16 H+3 \beta) A_1-\frac{1}{62} (256 A_2+48 \alpha-3 \gamma) A_2\\ & \displaystyle +\frac{\beta}{8} B_2+\frac{\beta}{32} H+\frac{\delta}{32} F-\frac{1}{8192} (256 \alpha^2+16 \beta^2+\gamma^2) +\frac{\alpha \gamma}{256}
   \end{array}
    \]
\[
\{\{B_1,B_2\},B_{2}\}=-\frac{1}{128} (256 A_2+16 \alpha+\gamma) A_1-\frac{\beta}{8} A_2-\frac{\gamma}{8} B_2-\frac{\delta}{64} B_1+\frac{\beta \gamma}{2048}-\frac{\alpha \beta }{128}
\]
\[
\begin{array}{cl}
        \{ F,\{F,A_{2}\}\}= & \displaystyle -\frac{1}{64} (384 A_1+1536 B_2+128 H+16 \beta+\delta) A_2+\frac{1}{64} (3\gamma-16 \alpha) H-\\ & -\frac{\gamma}{4} F- \displaystyle -\frac{1}{32} (128 B_1+16 \alpha +\gamma ) A_1-\frac{1}{32} (8 \beta+\delta) B_1- \\ & -\frac{1}{16} (48 \alpha+3 \gamma) B_2-\frac{\beta \gamma}{1024}-\displaystyle -\frac{\alpha \beta}{64}
        \end{array}
\]
\[
        \begin{array}{cl}
        \{ \{F,A_{2}\},A_{2}\}= & \displaystyle -\frac{1}{64} (128 A_1+32 \beta-\delta) A_1+\frac{\gamma}{8} A_2+\\
        &+\frac{3 \delta}{16} B_2+\frac{\delta}{64} H-\frac{1}{1024} (8 \beta^2-\gamma^2)+\frac{\alpha \gamma}{64}
        \end{array}
\]
The second algebra which expand in terms of $C_{1},C_{2},D, L$  with coefficients any linear combination of integrals $A_{1},A_{2},B_{2},H$
\[
\{ C_{1},C_{2} \}=0, \; \{C_{1},D \}=-\frac{\gamma}{8} C_{2}-\frac{\delta}{64} L, \; \{ C_{2},D \}=\frac{\gamma}{16} C_{2}+\frac{\delta}{128} L
\]
\[
\{ C_{1}, L \}=-\frac{\delta}{8} C_{2} \; , \{ C_{2}, L \}=\frac{\delta}{16} C_{2}
\]
\[
\begin{array}{l}
\displaystyle \{ D, L \}= \Bigl (  -\frac{A_{1} \delta ^2}{2 (16 \alpha -\gamma ) \gamma }+\frac{2
B_{2} \delta ^2}{(16 \alpha -\gamma ) \gamma }+\frac{H \delta
   ^2}{2 (16 \alpha -\gamma ) \gamma }+\frac{4 A_{2} \delta }{16
   \alpha -\gamma }+\frac{4 A_{1} \beta  \delta }{(16 \alpha -\gamma
   ) \gamma }-\\ \displaystyle -\frac{16 B_{2} \beta  \delta }{(16 \alpha -\gamma )
   \gamma }-\frac{4 H \beta  \delta }{(16 \alpha -\gamma ) \gamma
   }-\frac{4 A_{1} \gamma }{16 \alpha -\gamma }+\frac{64 A_{1}
   \alpha }{16 \alpha -\gamma }-\frac{32 A_{2} \beta }{16 \alpha
   -\gamma }\Bigr ) C_{2}+ \\ \displaystyle
   \Bigl (  -\frac{2 A_{1} \delta ^2}{(16 \alpha -\gamma ) \gamma }-\frac{\gamma
    \delta }{16 (16 \alpha -\gamma )}-\frac{\alpha  \delta }{16 \alpha
   -\gamma }+\frac{16 A_{1} \beta  \delta }{(16 \alpha -\gamma )
   \gamma }+\frac{\beta  \gamma }{16 \alpha -\gamma }  \Bigr ) D+ \\ \displaystyle + \Bigl ( -\frac{\beta ^2}{2 (16 \alpha -\gamma )}+\frac{\delta  \beta }{16 (16
   \alpha -\gamma )}-\frac{4 A_{2} \delta  \beta }{(16 \alpha
   -\gamma ) \gamma }-\\-\dfrac{8 A_{1} \beta }{16 \alpha -\gamma
   }-\dfrac{\gamma ^2}{16 (16 \alpha -\gamma )}+
   +\dfrac{A_{2} \delta
   ^2}{2 (16 \alpha -\gamma ) \gamma }+ \displaystyle  +\frac{\alpha  \gamma }{16 \alpha
   -\gamma }+\frac{A_{1} \delta }{16 \alpha -\gamma }   \Bigr ) L
\end{array}
\]
The relation between $C_{1},C_{2}$ is:
\[
C_{1}+2 C_{2}=0
\]

\end{itemize}

\section{Conclusions}\label{sec10}

The three dimensional non degenerate systems of  Kalnins, Kress and Miller \cite{KalKrMi07} satisfy a parafermionic like quadratic Poisson algebra.
All the algebras have at least a two-dimensional like parafermionic quadratic Poisson subalgebra. All the systems with one exception the have at least two subalgebras forming a special "$\Pi"$ structure. Each subalgebra coreesponds to classical superintegrable system possessing two Hamiltonians.

There is no results yet about the quantum  superintegrable systems as also a a compact general classification theory for three dimentional superintegrable potentials. The structure of the corresponding Poisson algebras for the degenarate systems is under investigation.

\end{document}